\documentclass[aps,prl,reprint,groupedaddress]{revtex4-1}
\usepackage{graphicx} % Include figure files
\usepackage{amsmath}
\usepackage{dcolumn} % Align table columns on decimal point
\usepackage{bm}% bold math
\usepackage{hyperref}% add hypertext capabilities

\begin{document}
\title{Exploring the neutron-star matter properties via the deformed nuclear reactions}

\author{Yuan-Qing Guo$^{1}$}
\author{Ya-Peng Zhang$^{2}$}
\author{Zhao-Qing Feng$^{1,2}$}
\email{Corresponding author: fengzhq@scut.edu.cn}

\affiliation{
$^{1}$School of Physics and Optoelectronics, South China University of Technology, Guangzhou 510640, China  \\
$^{2}$State Key Laboratory of Heavy Ion Science and Technology, Institute of Modern Physics, Chinese Academy of Sciences, Lanzhou 730000, China
}

\date{\today}

\begin{abstract}

Within the framework of Lanzhou quantum molecular dynamics transport model, the correlation of initial deformation and isospin diffusion is systematically investigated in collisions of $^{238}$U + $^{238}$U. The impacts of the collision centrality, symmetry energy and initial configuration on the collective flows, neutron/proton and $\pi^{-}/\pi^{+}$ ratios have been systematically investigated. It is found that the broader neutron-rich region is formed in the body-body collisions in comparison with the ones in the tip-tip collisions. The neutron-star matter might be created in the density region of 0.2-0.5 $\rho_{0}$ (the normal nuclear density $\rho_{0}$=0.165 fm$^{-3}$) formed in the $^{238}$U + $^{238}$U reaction at the incident energy of 500 MeV/nucleon. The elliptic flows of protons are related to the incident energy, collision centrality, symmetry energy and collision orientation. The hard symmetry energy enables the larger free neutron/proton ratio at the beam energy of 500 MeV/nucleon. However, the neutron/proton and $\pi^{-}/\pi^{+}$ ratios in the density regime of $1.2\leq \rho/\rho_{0}\leq1.8$ are enhanced by the soft symmetry energy with the slope parameter of L=42 MeV.

\end{abstract}

\maketitle

\section{I. Introduction}
The neutron-skin thickness of neutron-rich nuclide is correlated to the symmetry energy and neutron-star matter properties, which was proposed to be measured by the parity violating in electron-nucleus scattering \cite{Ho01a,Ho01b}. The isospin asymmetry of finite nucleus is far away from the ones of neutron-star matter and also deviated from the finite nucleus properties for extracting the neutron-star structure, i.e., the mass-radius relation, pasta phase transition etc. The nuclear collisions via the deformed nuclei enables very neutron-rich zone close to the neutron-star crust with the extremely isospin asymmetry nuclear matter. The nuclear shape configuration has been extensively investigated by the energy-density functional theories, i.e., the ground-state deformation, shape coexistence, triaxiality deformation, shape transition etc \cite{Mac77,Cwi05,Hey11}. The nuclear shape at ground state predominantly assumes an ellipsoidal shape (quadrupole deformation) in the heavy-mass region. Beside the electric or magnetic multipole $\gamma-$transition, the deformation coefficient of a atomic nucleus might be measured by the electron-nucleus scattering or high-energy heavy-ion collisions \cite{Hof56,Sud17,Gia20,Gia21,Jia22,Zha22,Star24}. Recently, it has been investigated that imaging the shape of atomic nuclei by analyzing the collective flows of emitted particles in relativistic heavy-ion collisions. The expansion of the quark-gluon plasma exhibits the anisotropy and is sensitive to the initial shape configuration due to the short reaction time.

Furthermore, intermediate energy heavy-ion collisions provide a unique way for exploring the dense nuclear matter properties in terrestrial laboratories, which can be used effectively to study the properties of nuclear matter and to extract information of the symmetry energy for the nuclear equation of state (EOS), which is expressed with the energy per nucleon as E($\rho,\delta$) = E($\rho,\delta=0$) + $E_{sym}(\rho)\delta^2$ + $\mathcal{O}(\delta^4)$ in terms of baryon density $\rho$ = $\rho_n$+$\rho_p$ and relative neutron excess $\delta$ = ($\rho_n - \rho_p$)/($\rho_n + \rho_p$), where $E_{sym}(\rho)$ is the symmetry energy \cite{Bom91,Li02,Li08}. The symmetry energy at supersaturated densities is an important component of the equation of state of nuclear matter,  which is of significance to the understanding of properties of nuclear matter, the mass-radius relation of neutron stars, and the hyperon production etc \cite{Bar05,Ste05,Lat04,Zha20,Fen23}. The dense nuclear matter might be created via the nuclear collisions in laboratories, such as the Cooling Storage Ring (CSR) and the High Intensity Accelerator Facility (HIAF) in China \cite{Yan13,Ch19}, Radioactive Isotope Beam Facility (RIBF) in Japan \cite{Sak18}, Rare Isotope Science Project in Korea (RAON) \cite{Tsh13}, and Facility for Rare Isotope Beams (FRIB) in the USA \cite{Ost19}.

The reactions with the deformed neutron-rich nuclei, e.g., the collisions of $^{238}$U + $^{238}$U, will manifest the different isospin diffusion with the collision orientation. The deformation effect, i.e., the tip-tip (short-axis to short-axis) and body-body (long-axis to long-axis) collisions, is related to the initial shape configuration. In heavy-ion collisions, the nuclear stopping effect, collective flows, particle production, cluster formation etc, are influenced by the collision centrality and orientation \cite{Li00,Luo07,Cao10,Xu12,Das17,Fan23,Liu23}. In this work, the deformation and isospin effects in the reaction of $^{238}$U + $^{238}$U are to be systematically investigated at the incident energy of 500 MeV/nucleon (CSR energy regime) by using the Lanzhou quantum molecular dynamics (LQMD) transport model.

\section{II. Model description }

In the LQMD transport model, the production of resonances, hyperons and mesons is coupled to the channels in the meson-baryon and baryon-baryon reactions, which has been investigated for the nuclear dynamics in heavy-ion collisions and hadron induced reactions \cite{Fen18,Fen11,Fen12,Wei24}. The temporal evolutions of nucleons and nucleonic resonances are described by Hamilton's equations of motion under the self-consistently generated two-body and three-body potentials with the Skyrme effective interaction. The symmetry energy is composed of three parts, namely, the kinetic energy from Fermi motion, the local density-dependent interaction, and the momentum-dependent potential, which reads as
\begin{equation}
E_{sym}(\rho)=\frac{1}{3}\frac{\hbar^{2} }{2m} (\frac{3}{2}\pi ^{2}\rho  )^{2/3} +E^{loc}_{sym}(\rho)+E^{mom}_{sym}(\rho).
\end{equation}
with $m$ being the nucleon mass. The stiffness of symmetry energy is adjusted by
\begin{equation}
E_{sym}^{loc}\left ( \rho  \right ) = \frac{1}{2} C_{sym}\left ( \rho /\rho _{0}  \right )^{\gamma _{s} }.
\end{equation}
The parameter $C_{sym}$ is 52.5 MeV, and the stiffness parameter $\gamma_s$ is adjusted for getting the density dependence of symmetry energy, e.g., the values of 0.3, 1, and 2 being the soft, linear, and hard symmetry energy, corresponding to the slope parameters $[L(\rho _{0}) = 3\rho_{0} dE_{sym}(\rho)/d\rho|_{\rho=\rho_{0}}]$ of 42, 82, and 139 MeV, respectively. The stiffness of symmetry energy might be constrained by the isospin observables produced in different density region in heavy-ion collisions.

The reaction channels for the pion production contributed from the direct process and the resonance decay, such as $\Delta$(1232), $N^{\ast}$(1440), $N^{\ast}$(1535) etc, are included in the model as follows \cite{Fen18}
\begin{eqnarray}
&& NN \leftrightarrow N\triangle, \quad  NN \leftrightarrow NN^{\ast}, \quad  NN
\leftrightarrow \triangle\triangle,  \nonumber \\
&& \Delta \leftrightarrow N\pi,  N^{\ast} \leftrightarrow N\pi,  NN \leftrightarrow NN\pi  (\texttt{s-state}),
\end{eqnarray}
Here the nucleon-nucleon (resonance) and pion-nucleon collisions are treated as the stochastic and isotropic scattering in the temporal evolution of reaction system. The momentum-dependent decay widths are implemented into the model for the resonances of $\Delta$(1232) and $N^{\ast}$(1440) and the elementary cross sections are taken from the parameterized formula calculated by the one-boson exchange model \cite{Hub94}. The constant width of $\Gamma$=150 MeV for the $N^{\ast}$(1535) decay is used in the calculation. The elastic scatterings in nucleon-nucleon, nucleon-resonance ($NR\rightarrow NR$) and resonance-resonance ($RR\rightarrow RR$) collisions and inelastic collisions of nucleon-resonance ($NR\rightarrow NN$, $NR\rightarrow NR^{\prime}$) and resonance-resonance channels ($RR\rightarrow NN$, $RR\rightarrow NR$, $RR\rightarrow RR^{\prime}$, $R$ and $R^{\prime}$ being different resonances), have been included in the model. The direct process $ NN \leftrightarrow NN\pi  (\texttt{s-state})$ roughly contributes 15$\%$ pion yields and the cross sections are taken as the same with the Giessen Boltzmann-Uehling-Uhlenbeck (GiBUU) transport model \cite{Bus12}.

The transportation of pions in nuclear medium is also described by the Hamiltonian equation of motion as
\begin{equation}
 H_{M} = \sum_{i=1}^{N_{M} }  \left[ V_{i}^{Coul} +  \omega \left (\textbf{p}_{i},\rho _{i}   \right)  \right].
\end{equation}
The Coulomb interaction is given by
\begin{equation}
    V_{i}^{Coul}= \sum_{j=1}^{N_{B} } \frac{e_{i}e_{j} }{r_{ij} }
\end{equation}
with $r_{ij}=|\textbf{r}_{i} - \textbf{r}_{j}|$. Here the $N_M$ and $N_B$ are the total numbers of mesons and baryons including charged resonances, respectively.

The pion energy in the nuclear medium is composed of the isoscalar and isovector contributions as
\begin{equation}
    \omega _{\pi }  \left( \textbf{p}_{i} ,\rho _{i}  \right) = \omega _{isoscalar} \left(\textbf{p}_{i} ,\rho _{i}  \right) + C_{\pi }\tau _{z}\delta \left (\rho /\rho _{0}  \right )^{\gamma _{\pi } }.
\end{equation}
Here the isovector coefficient $C_{\pi}$ = $\rho_{0}\hbar^{3}/(4f^{2}_{\pi})$ = 36 MeV, isospin asymmetry $\delta$=($\rho_n - \rho_p$)/($\rho_n + \rho_p$ ) and isospin splitting parameter $\gamma_{\pi}$=2. The isospin quantities are set to be $\tau _{z}$= -1, 0, and 1 for $\pi^{+}$, $\pi^{0}$, and $\pi^{-}$, respectively \cite{Fen15}. The isoscalar part $\omega _{isoscalar} $ is estimated by the $\Delta$-hole model \cite{Bro75,Fri81}.

 The energy balance in the decay of resonances and reabsorption of pions in nuclear medium is satisfied by the relation $R \leftrightarrow N\pi $ ($R$ being the resonance) as
 \begin{eqnarray}
&& \sqrt{m_{R}^{2}+\textbf{p}_{R}^{2}} + U_{R}(\rho,\delta,\textbf{p}_{R})  = \sqrt{m_{N}^{2} + \left (\textbf{p}_{R} - \textbf{p}_{\pi } \right )^{2} }         \nonumber \\
&& + U_{N}(\rho,\delta,\textbf{p})  + \omega _{\pi} \left (\textbf{p}_{\pi },\rho  \right)+V_{\pi N}^{Coul}.
 \end{eqnarray}
The $\textbf{p}_{R}$ and $\textbf{p}_{\pi}$ are the momenta of resonance and pion, respectively. The term $V_{\pi N}^{Coul}$ has the contribution only for the charged pair channels of $\triangle^{0}\leftrightarrow  \pi^{-}+p$ and $\triangle^{++}\leftrightarrow  \pi^{+}+p$, and no effect for the channels associated with the $\pi^{0}$ and neutron production. The optical potential can be evaluated from the in-medium energy $V_{\pi}^{opt}(\textbf{p},\rho) = \omega_{\pi} (\textbf{p},\rho) - (m_{\pi}^{2}+\textbf{p}^{2})^{1/2}$. The $U_{R}$ and $U_{N}$ are the singe-particle potentials for resonance and nucleon, respectively. For example, the $\Delta$(1232) optical potential is calculated via the nucleon optical potential by \cite{LiuH23}
 \begin{eqnarray}
  && U_{\Delta ^{- } }  =U_{n}, \quad   U_{\Delta ^{++ } }  =U_{p},  \quad
  U_{\Delta ^{+ }} = \frac{1}{3} U_{n} + \frac{2}{3}U_{p},      \nonumber \\
  &&  U_{\Delta ^{0 } }  =\frac{1}{3} U_{p}+  \frac{2}{3}U_{n},
\end{eqnarray}
 where the $U_{n}$ and $U_{p}$ are the single-particle potentials for neutron and proton, respectively.

\begin{figure}
 \includegraphics{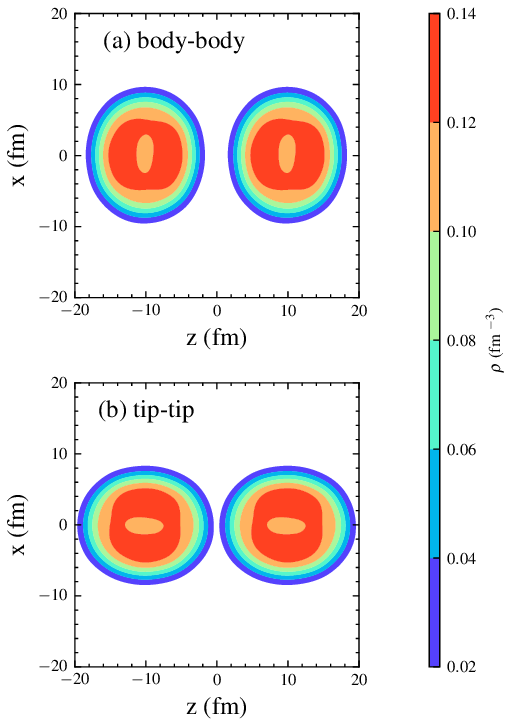}
 \caption{ The density profiles in the (a) body-body and (b) tip-tip orientations in the reaction of $^{238}$U + $^{238}$U. }
\label{fig:1}
\end{figure}

In the initialization, the root-mean-square (rms) radii is sampled with the quadrupole deformation and rechecked with the ones calculated by the well-known Skyrme-Hartree-Fork method as follows
\begin{equation}
    R_{n,p} = R_{0}(1 + \beta_{2}\sqrt{\frac{5}{4\pi}}\frac{3\cos^2\theta - 1 }{2})
\end{equation}
for proton and neutron profiles, respectively. The quadrupole deformation parameter ${\beta_{2}}$, e.g., 0.215 for $^{238}$U, the relation of nuclear radius and mass number is given by $R_{0} = 1.28A^{1/3} + 0.8/A^{1/3} - 0.76$ fm. The long and short axes of ellipsoid configuration are related to the deformation as
\begin{equation}
    \begin{aligned}
    R_{\text{long}}   = R_{0}(1 + \beta_{2}\sqrt{\frac{5}{4\pi}}),   \\
    R_{\text{short}} = R_{0}(1 - \frac{\beta_{2}}{2}\sqrt{\frac{5}{4\pi}}).
    \end{aligned}
\end{equation}
The momentum of each nucleon is uniformly sampled within a Fermi sphere of radius $p_{n,p}(\mathbf{r}) = \hbar (3\pi^2\rho_{n,p}(\mathbf{r}))$. The ground-state binding energy and rms radii in the initialization nucleus are reexamined by the available experimental data. Shown in Fig. \ref{fig:1} is the initial density profiles in the body-body and tip-tip orientations for the reaction of $^{238}$U + $^{238}$U in the $x-z$ plane. The neutron-skin thickness with the nuclear density below 0.08 fm$^{-3}$ is obvious and influences the isospin diffusion in different collision orientation, which leads to the formation of different reaction zone of neutron-rich matter. 

%%%%%%%%%%%%%%%%%%%%%%%%%%%%%%%%%%%% figure 1 %%%%%%%%%%%%%%%%%%%%%%
%%%%%%%%%%%%%%%%%%%%%%%%%%%%%%%%%%%%%%%%%%%%%%%%%%%%%%%%%%%%%%%

\section{III. Results and discussion}
The anisotropic emission of particles is associated with the dense nuclear matter properties. The collective flows in heavy-ion collisions manifest the anisotropic distribution in phase space and reply on the beam energy, collision centrality, initial shape configuration etc. The azimuthal distribution of particle production is expressed in Fourier series form as \cite{Ram99}
\begin{equation}
\frac{dN}{d\phi} = N_0 \left(1 + \sum_{n} 2v_n \cos(n\phi)\right),
\end{equation}
where the second-order coefficient is defined as the elliptic flow and given by
\begin{equation}
v_2(P_{t},y) = \left \langle(p^{2}_{x}-p^{2}_{y})/p^{2}_{t}\right \rangle =\left \langle cos(2\phi)\right \rangle.
\end{equation}
The in-plane $(v_{2}>0)$ and out-of-plane $(v_{2}<0)$ particle emissions are related to the transverse momentum
$P_t=\sqrt{p_x^2+p_y^2}$ and the longitudinal rapidity $y=\frac{1}{2}\ln{\frac{E+p_z}{E-p_z}}$ with the total energy $E$, respectively.

%%%%%%%%%%%%%%%%%%%%%%%%%%%%%%%%%%%% figure 2 %%%%%%%%%%%%%%%%%%%%
\begin{figure}
    \includegraphics[height=4cm,width=9cm]{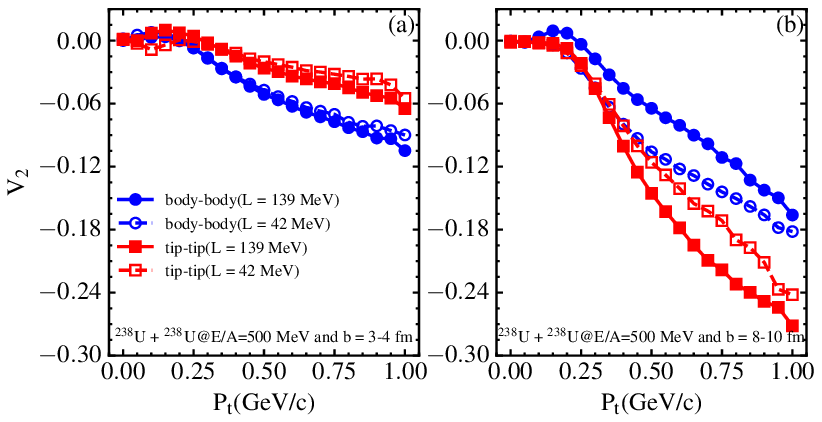}
    \caption{Transverse momentum distribution of the elliptic flows for free protons in collisions of $^{238}$U + $^{238}$U at the incident energy of 500 MeV/nucleon with different collision orientations and symmetry energy. }
    \label{fig.2}
\end{figure}
%%%%%%%%%%%%%%%%%%%%%%%%%%%%%%%%%%%%%%%%%%%%%%%%%%%%%%%%%%%%%

%%%%%%%%%%%%%%%%%%%%%%%%%%%%%%%%%%%% figure 3 %%%%%%%%%%%%%%%%%%%%
\begin{figure*}
 \includegraphics[height=6cm,width=18cm]{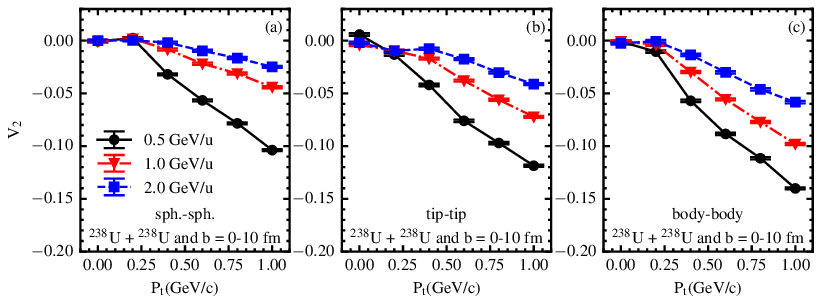}
   \caption{The transverse momentum dependence of the proton elliptic flows at the incident energies of 0.5, 1.0 and 2.0 GeV/nucleon in $^{238}$U + $^{238}$U collisions at (a) the spherical-spherical (no deformation), (b) tip-tip and (c) body-body orientations, respectively.}
    \label{fig.3}
\end{figure*}
%%%%%%%%%%%%%%%%%%%%%%%%%%%%%%%%%%%%%%%%%%%%%%%%%%%%%%%%%%%%%

The isospin diffusion is caused by the difference of isospin density formed in heavy-ion collisions and influenced by the stiffness of symmetry energy and collision configuration. The anisotropic emission of particles is correlated to the initial configuration. Shown in Fig. 2 is a comparison of collision orientation (body-body and tip-tip) and different symmetry energy in the reaction of $^{238}$U + $^{238}$U at an incident energy of 500 MeV/nucleon. The system is being planned for investigating the dense matter properties via the light charged particles at the Cooling-storage- ring External target Experiment (CEE) in Lanzhou. It is obvious that the anisotropic emission is pronounced in the peripheral collisions owing to the geometrical asymmetry. More collision nucleons appear in the tip-tip orientation for the peripheral collisions and result in the larger elliptic flows. However, the body-body orientation is favorable for the anisotropic proton emission in semi-central collisions. On the other hand, the hard symmetry energy with L=139 MeV enables more repulsive for the proton emission in the tip-tip collisions. Overall, the anisotropic proton emission relies on the reaction zone in the nucleus-nucleus collisions, which is different in the tip-tip and body-body collisions. The influence of the symmetry energy on the proton emission is sensitive to the compression region formed in nuclear collisions. The hard symmetry energy (L=139 MeV) leads to the more repulsive for protons in the tip-tip collisions. The initial shape configuration has more pronounced effect on the proton anisotropic emission than the
ones of symmetry energy. The elliptic flows depend on the incident energy in heavy-ion collisions as shown in Fig. 3.
The comparison of 0.5, 1.0 and 2.0 GeV/nucleon in collisions of $^{238}$U + $^{238}$U is presented at the spherical-spherical, tip-tip and body-body orientations, respectively. The beam energy dependence of elliptic flow manifests the competition of nucleon-nucleon collision and mean-field potential in the dense nuclear matter. It has been observed that the elliptic flow monotonically decreases with the beam energy until to $\sqrt{s_{NN}}=2.76$ GeV \cite{Ali10,Zh25}. The negative elliptic flows are mainly caused from the repulsive interaction in the dense matter. It is obvious that the flow $v_{2}$ goes up with increasing the incident energy at the different initial configuration.

%%%%%%%%%%%%%%%%%%%%%%%%%%%%%%%%%%%% figure 4 %%%%%%%%%%%%%%%%%%%%
\begin{figure}
    \includegraphics[height=4cm,width=8cm]{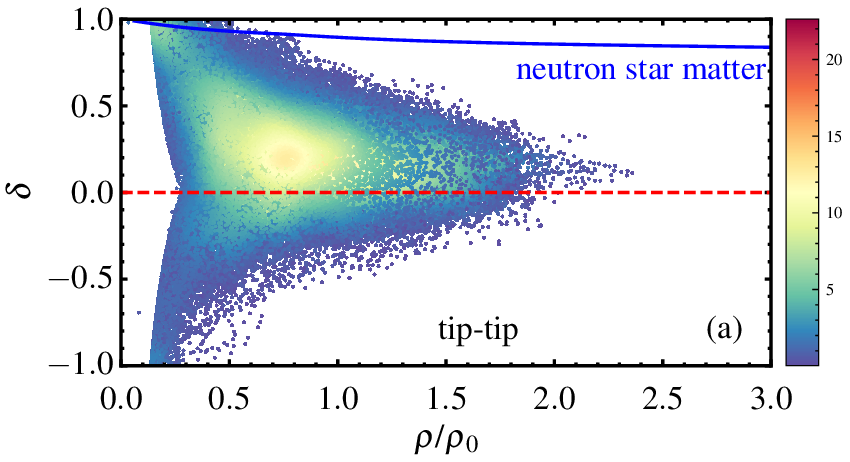}
    \includegraphics[height=4cm,width=8cm]{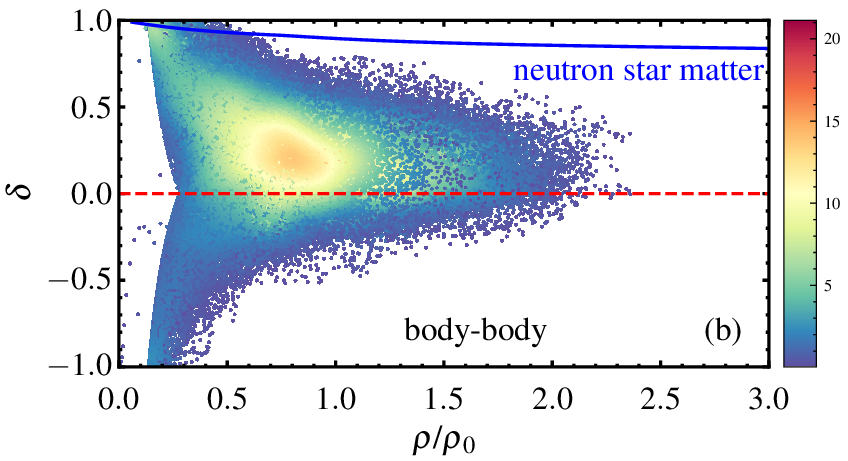}
    \caption{ The isospin asymmetry as a function of the nuclear density in the temporal evolution of $^{238}$U + $^{238}$U at 500 MeV/nucleon with the peripheral collisions. }
    \label{fig.4}
\end{figure}
%%%%%%%%%%%%%%%%%%%%%%%%%%%%%%%%%%%%%%%%%%%%%%%%%%%%%%%%%%%%%

The nuclear matter formed in heavy-ion collisions is the less isospin asymmetry ($\delta = (\rho_n - \rho_p)/(\rho_n + \rho_p)$) in comparison with the neutron-star structure. The neutron-rich matter and observables in nuclear dynamics are particular important for investigating the neutron-star properties in terrestrial laboratories. Besides the density dependence of symmetry energy, the cluster formation in the neutron-rich matter, the short-range correlation, hyperon appearance in neutron stars etc, influence the mass-radius relation. The correlation of the isospin asymmetry and nuclear density in the reaction of $^{238}$U + $^{238}$U at 500 MeV/nucleon is calculated as shown in Fig. 4. It is obvious that the broad neutron-rich matter is created in the body-body collisions. The density regime of 0.2-0.6$\rho_{0}$ in the reaction is close to the isospin asymmetry of neutron-star matter. The corresponding observables emitted in the neutron-rich regime are particularly important for constraining the symmetry energy.

%%%%%%%%%%%%%%%%%%%%%%%%%%%%%%%%%%%% figure 5 %%%%%%%%%%%%%%%%%%%%
\begin{figure}
\includegraphics[height=6cm,width=6cm]{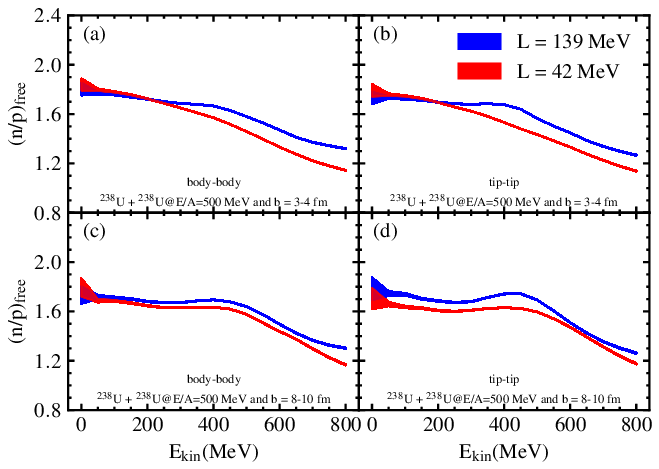}
\caption{The kinetic energy spectra of free neutron/proton ratio produced in collisions of $^{238}$U + $^{238}$U at 500 MeV/nucleon with the (a) body-body orientation and centrality of 3-4 fm, (b) tip-tip orientation and centrality of 8-10 fm, (c) body-body orientation and centrality of 3-4 fm, (d) tip-tip orientation and centrality of 8-10 fm, respectively. }
    \label{fig.5}
\end{figure}
%%%%%%%%%%%%%%%%%%%%%%%%%%%%%%%%%%%%%%%%%%%%%%%%%%%%%%%%%%%%%

The observables produced from the neutron-rich region is of importance for probing the neutron-star matter properties. The symmetry potential and the gradient of isospin density influence the phase-space evolution of the experimental observables. Shown in Fig. 5 is the kinetic energy spectra of free neutron/proton (N/Z) ratio produced in collisions of $^{238}$U + $^{238}$U at 500 MeV/nucleon with the different initial configuration and collision centrality. It is obvious that the hard symmetry energy (L=139 MeV) leads to the larger N/Z ratio because of the more repulsive interaction for neutrons in the neutron-rich matter, in particular in the semi-central collisions (3-4 fm) at the kinetic energies above 300 MeV. It should be noticed that the energy spectra of $^{3}$H/$^{3}$He ratio
is also a sensitive probe for extracting the high-density symmetry energy. The kinetic approach for cluster production in heavy-ion collisions is necessary for investigating the neutron-rich matter properties, in which the binding energies of clusters, Mott effect, cluster transportation in nuclear medium etc, need to be taken into account.

%%%%%%%%%%%%%%%%%%%%%%%%%%%%%%%%%%%% figure 6 %%%%%%%%%%%%%%%%%%%%
\begin{figure}
    \includegraphics[height=4cm,width=8cm]{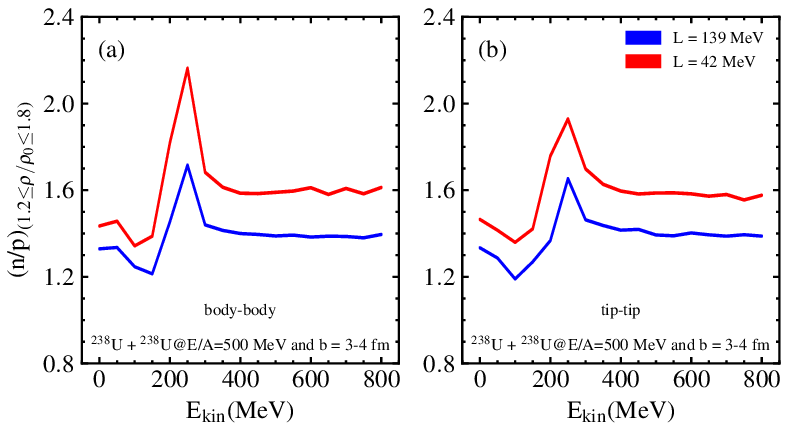}
    \includegraphics[height=4cm,width=8cm]{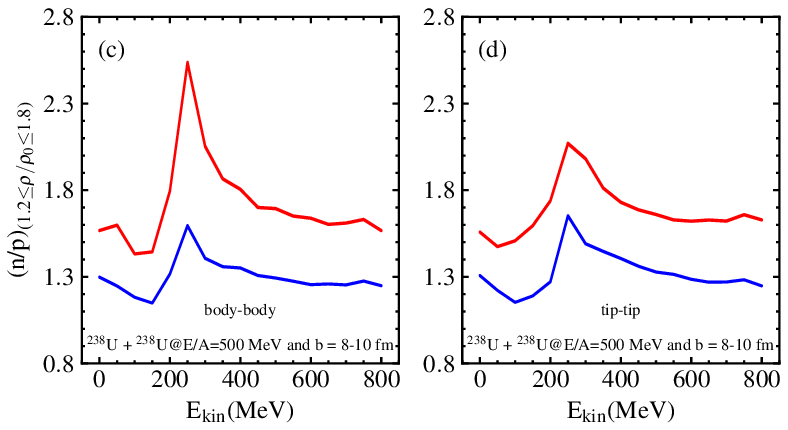}
    \caption{The neutron/proton ratio at the suprasaturation densities 1.2$\leq\rho/\rho_{0}\leq$1.8 with the temporal evolution in collisions of $^{238}$U + $^{238}$U at the beam energy of 500 MeV/nucleon and with different collision centrality and orientation. }
    \label{fig.6}
\end{figure}
%%%%%%%%%%%%%%%%%%%%%%%%%%%%%%%%%%%%%%%%%%%%%%%%%%%%%%%%%%%%%

%%%%%%%%%%%%%%%%%%%%%%%%%%%%%%%%%%%% figure 7 %%%%%%%%%%%%%%%%%%%%
\begin{figure}
    \centering
    \includegraphics[height=4cm,width=8cm]{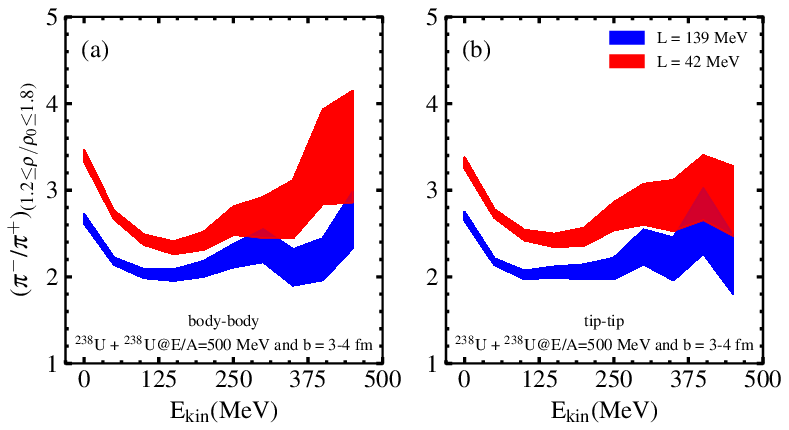}
    \includegraphics[height=4cm,width=8cm]{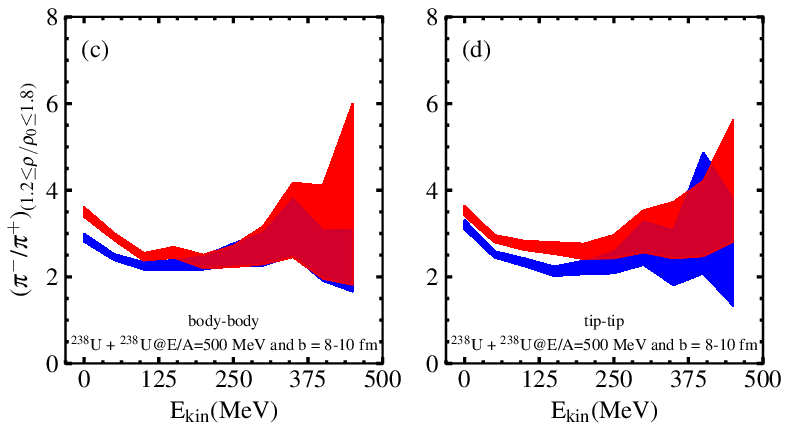}
    \caption{The kinetic energy spectra of $\pi^{-}/\pi^{+}$ ratio in the density domain of 1.2$\leq\rho/\rho_{0}\leq$1.8 in  collisions of $^{238}$U + $^{238}$U at 500\emph{A} MeV with the (a) (c) body-body and (b) (d) tip-tip orientations, respectively.}
 \label{fig.7}
\end{figure}
%%%%%%%%%%%%%%%%%%%%%%%%%%%%%%%%%%%%%%%%%%%%%%%%%%%%%%%%%%%%%

The experimental observables produced in the high-density regime in heavy-ion collisions are available for probing the inner structure of neutron stars, e.g., the isospin ratios, $\pi^{-}/\pi^{+}$, K$^{0}$/K$^{+}$, $\Sigma^{-}/\Sigma^{+}$, $\Xi^{-}/\Xi^{0}$ etc \cite{Fen24}. The hadrons are modified in the dense nuclear medium, i.e., the production and transportation, effective mass, rescattering by the surrounding nucleons etc. The kinetic energy spectra of neutron/proton ratio in the density region of 1.2-1.8 $\rho/\rho_0$ in collisions of $^{238}$U + $^{238}$U are shown in Fig. \ref{fig.6}. The soft symmetry energy with L=42 MeV enlarges the n/p ratio and leads to the formation of more neutron-rich matter. On the other hand, the broader neutron-rich matter is formed in the body-body collisions. The bump structure is caused from the appearance of pion production at the threshold energy of 280 MeV. Similarly, the pion spectra are also calculated as shown in Fig. 7. It is obvious that the symmetry energy effect is pronounced for the $\pi^{-}/\pi^{+}$ ratio in the semi-central collisions. The initial configuration weakly influences the pion spectra. The pion production in heavy-ion collisions is complicated and associated with the pion-nucleon potential, the in-medium properties of $\Delta(1232)$ etc. The measurements of the production in heavy-ion collisions at the CEE spectrometer are expected for exploring the pion-nucleon interaction in dense matter, high-density symmetry energy, in-medium properties of $\Delta(1232)$ (decay width, rescattering, interaction with nucleons) etc.

\section{IV. Conclusions}
In summary, we have investigated the correlation of the initial quadrupole deformation effect and isospin diffusion in the reaction of $^{238}$U + $^{238}$U within the LQMD transport model. The system is planned for the upcoming experiments at HIRFL-CSR with the beam energy of 0.5 GeV/nucleon. The neutron-rich matter formed in the reaction in the density regime of 0.2-0.6$\rho_{0}$ is close to the isospin asymmetry of neutron-star matter and related to the collision orientation. The broader neutron-rich matter is created in the body-body collisions. The squeeze-out protons are pronounced at the incident energy of 0.5 GeV/nucleon in comparison with the ones at 1.0 GeV/nucleon and 2.0 GeV/nucleon. The free n/p ratio is enhanced by the hard symmetry energy, in particular in the central collisions. However, the soft symmetry energy with the slope parameter of 42 MeV leads to the larger n/p and $\pi^{-}/\pi^{+}$ ratios in the nuclear matter at the densities of $1.2\leq \rho/\rho_{0}\leq1.8$.

\textbf{Acknowledgements}
This work was supported by the National Natural Science Foundation of China (Projects No. 12175072, No. 12475133 and No. 12311540139).

\bibliographystyle{unsrt}
\bibliography{reference}

\end{document}